# Atomistic mechanism and interface-structure-energetics of van der Waals epitaxy demonstrated by layered α-$MoO_3$ growth on mica

Faezeh A. F. Lahiji[1,2,3], Davide G. Sangiovanni[4], Biplab Paul[1,5], Justinas Palisaitis[1], Per O. Å. Persson[1], Arnaud le Febvrier[2], Ganpati Ramanath[1,2,6,7,*], Per Eklund[1, 2,7,*]

[1] *Inorganic Chemistry, Department of Chemistry-Ångström Laboratory, Uppsala University, Box 538, SE-751 21 Uppsala, Sweden*

[2]*Thin Film Physics Division, Department of Physics, Chemistry and Biology, (IFM), Linköping University, SE-58183 Linköping, Sweden*

[3]*Chair of Materials Chemistry, RWTH Aachen University, Kopernikusstraße 10, D-52074, Aachen, Germany*

[4]*Theoretical Physics Division, Department of Physics, Chemistry, and Biology (IFM) Linköping University, SE-581 83, Linköping, Sweden*

[5]*PLATIT AG, Eichholzstrasse 9, 2545 Selzach, Switzerland*

[6] *Materials Science & Engineering Department, Rensselaer Polytechnic Institute, Troy, NY 12180, USA*

[7]*Wallenberg Initiative Materials Science for Sustainability, Department of Chemistry-Ångström Laboratory, Uppsala University, Box 538, SE-751 21 Uppsala, Sweden*

***Corresponding authors:**

Prof. P. Eklund per.eklund@kemi.uu.se

Prof. G. Ramanath ganapr@rpi.edu ; ganpati.ramanath@kemi.uu.se

Unlike conventional epitaxy, van der Waals epitaxy (vdWE) allows nearly-stress-free growth of thick films with highly oriented crystals without dislocations even for large film-substrate lattice mismatches. Despite reports of vdWE in numerous materials systems, an atomistic understanding of film/substrate interface structure that explains and predicts vdWE has remained elusive. Here, we address this knowledge gap by unveiling atomistic interface mechanisms for vdWE of α-$MoO_3$(0k0) on mica(001). X-ray diffraction and electron microscopy reveal α-$MoO_3$(0k0) epilayers with large columnar crystals in three non-equivalent in-plane orientations. These results, together with negligible strain buildup in continuous epilayers, confirm vdWE. *Ab initio* computations showing interface energy minima for these orientations correlate with high cross-interface proximity between Mo atoms in α-$MoO_3$ and K in mica conducive for maximal vdW attraction. These atomistic insights on interface structure and energetics provide a crucial framework for predicting vdWE for different film/substrate combinations and designing of stress-free and/or standalone epitaxial films of layered materials such as $MoO_3$ on layered substrates such as f-mica.

**Background and context**

There is a great deal of interest in rationally realizing van der Waals epitaxy (vdWE), a mechanism characterized by atomic-level registry across a weakly-bonded film-substrate interface [1,2]. Weak interface bonding is conducive for stress-free crystal growth, and its release and transfer on to other substrates (e.g., amorphous) [3–5] or use as stand-alone films [6]. This is in contrast to conventional epitaxy with strong film-substrate interface bonding and lattice-match leads to strain buildup with increasing film thickness $t_{film}$ and subsequent relaxation by dislocation formation or roughening [7,8].

Since the discovery of vdWE in 1984 [9,10], vdWE has been claimed in numerous materials systems. The most intuitive examples of vdWE are seen in two-dimensional (2D) materials [1,11–17] like graphene [18] and hexagonal boron nitride [19], wherein the film-substrate atomic registry is mediated solely by vdW bonding. This attribute can be harnessed to facilitate vdWE by introducing graphene interlayers to grow highly oriented films on substrates with large mismatches in lattice parameter and symmetry [20].

The situation is more complex for substrates on which both conventional and vdW epitaxy are possible. A particularly important example is mica, a layered material commonly used for mechanically flexible thin films [21–25]. Mica has been shown to support vdWE for layered materials such as chalcogenides (e.g., $MoSe_2$, $MoS_2$, $Bi_2Te_3$)[5,10,19,26–28]. However, the current understanding of interface energetics and mechanisms of vdWE is inadequate to predict if a material will grow by vdWE or conventional epitaxy. As a result, vdWE is often *incorrectly presumed* (e.g., in nonlayered ZnO, GaN, AlN) simply because mica is layered, without establishing the conditions to indicate vdWE or even ignoring evidence to the contrary. In fact, for nonlayered materials such as transition metal oxides and nitrides [21,29,30], conventional epitaxy should be the default assumption for non-layered materials on mica, and any claim of vdWE must, as a minimum, provide compelling evidence. Necessary conditions for vdWE [3,3–6,18,19,21,21,31,32,32–34] are: a) in-plane and b) out-of-plane texturing, and c) nearly strain-free epilayers [35,36] independent of lattice mismatch or film thickness.

Here, we reveal new atomistic insights on interface registry and energetics underpinning vdWE of a layered material on a layered-material substrate, using the model case of $\alpha$-$MoO_3$ film on fluorophlogopite mica (f-mica). In particular, we show that proximal film-substrate interface atomic registry over long in-plane distances drives vdWE of $\alpha$-$MoO_3$ on f-mica. X-ray diffraction and electron microscopy reveal three non-equivalent in-plane domains of $\alpha$-$MoO_3$(0k0) crystals on f-mica(001). Continuous $\alpha$-$MoO_3$ epilayers are strain-free, as indicated by the fact that of the out-of-plane 0k0 spacings are independent on epilayer thickness. *Ab initio* computations reveal that the three domains correspond to vdW-bonding-driven energy minima informed by proximal cross-interface atomic correlations between Mo in $\alpha$-$MoO_3$ and K in f-mica. These insights will be invaluable to engineer stress-free vdW epilayers of a layered material such as $\alpha$-$MoO_3$ on layered substrates such as mica.

## Out-of-plane texture and stress evolution

X-ray diffractograms from molybdenum oxide films of thicknesses $2.5 \leq t_{film} \leq 40$ nm on f-mica (Fig. 1a) predominantly exhibit $\alpha$-$MoO_3$(0k0) reflections besides f-mica peaks, indicating strong out-of-plane (0k0) fiber texture. Magnified views of $\alpha$-$MoO_3$(060) peaks (Figs. 1b) show no overlap with the f-mica peaks [37]. Very thick films (e.g., $t_{film} = 160$ nm) show some loss of texture indicated by traces of $\alpha$-$MoO_3$(110) and (0 12 0) peaks at $2\theta = 23.08^\circ$ and $83.04^\circ$. $\alpha$-$MoO_3$ films grown on c-sapphire under similar conditions also exhibit fiber texture (Fig. S1a).

For $\alpha$-$MoO_3$ on f-mica, the ≈ 0.5% decrease in $\alpha$-$MoO_3$(060) spacing $d_{060}$ with increasing film thickness for $2.5 \leq t_{film} \leq 10$ nm (Fig. 1c) is attributed to decreased biaxial tensile strain as $\alpha$-$MoO_3$ crystal islands coalesce to form a continuous film. This view is supported by the 0k0 peak width decrease for $t_{film} \leq 10$ nm (Fig. 1d) indicating larger crystals and lower defects [38]. For $t_{film}$ >≈ 10 nm (≈8 unit cells), the much smaller (≤ 0.13%) decrease in $d_{060}$ and the 0k0 peak-width implies negligible strain build-up with increasing $t_{film}$, a defining feature of vdWE.

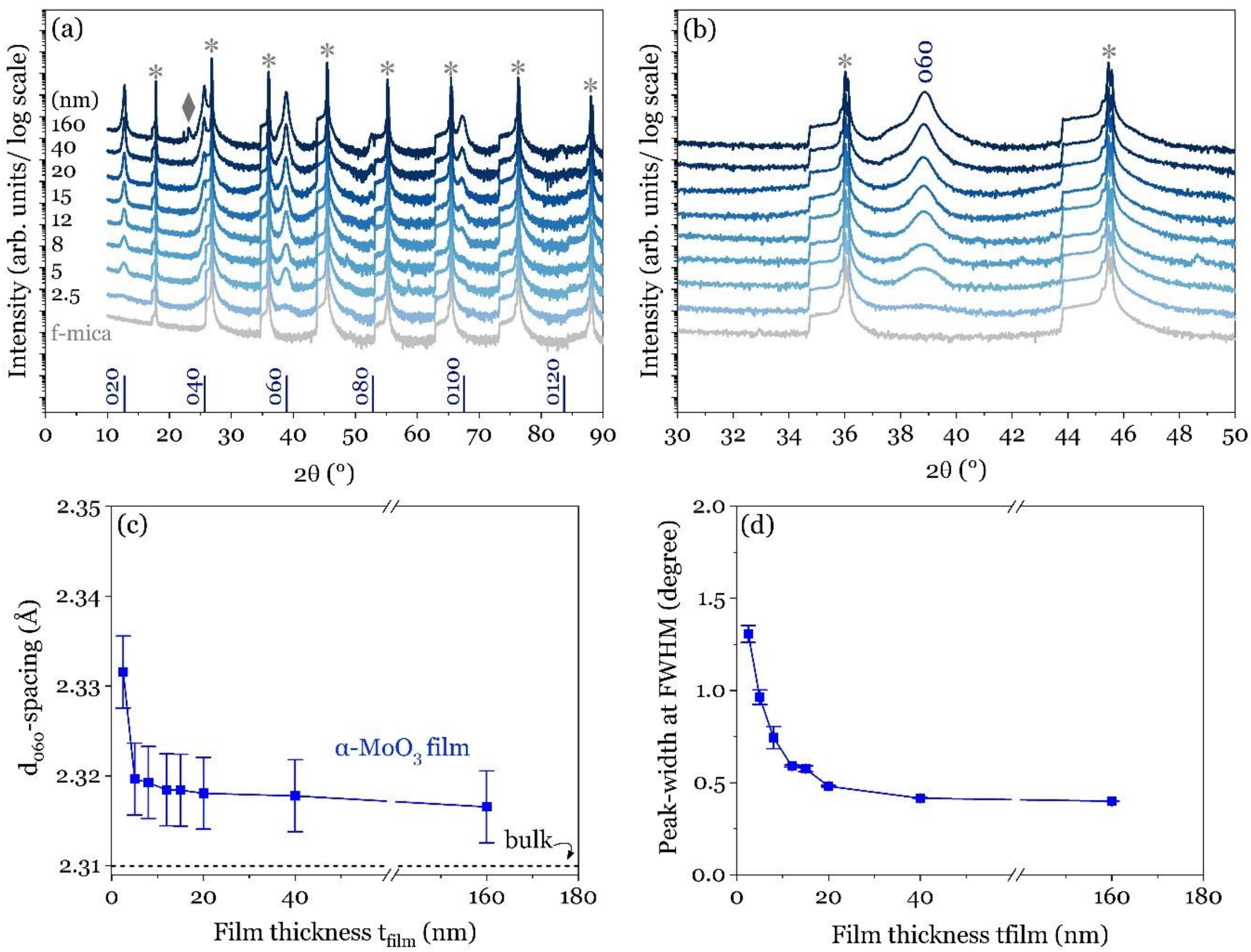

Fig. 1. X-ray diffractograms from $\alpha$-$MoO_3$ films with different thicknesses $t_{film}$ grown on f-mica (a), and magnified views of the $\alpha$-$MoO_3$(060) reflection (b), where "*" connote f-mica peaks. The weak diffraction peaks at $2\theta = 23.08^{\circ}$ corresponding to the 110 reflection from $\alpha$-$MoO_3$ (marked by a diamond) is < 1% of the $\alpha$-$MoO_3$(060) reflection intensity. $\alpha$-$MoO_3$(060) d-spacings (c) and full-width-half-maximum width of $\alpha$-$MoO_3$(060) reflections plotted versus $t_{film}$ (d).

The above observations are contrary to conventional epitaxy of $\alpha$-$MoO_3$ films on c-sapphire, wherein strong interface bonding leads to strain buildup with increasing film thickness and relaxation even for $t_{film} \approx 40$ nm, which corresponds to ≈ 30-unit cells (Fig. S1c). The narrower 0k0 peak width in α-$MoO_3$ films on f-mica (Fig. 1d), than on c-sapphire (Fig. S1d) indicates a higher coherence length in the α-$MoO_3$ crystals along out-of-plane $[010]_{\alpha\text{-}MoO3}$ growth direction. These results highlight the importance and impact of film/substrate bonding on the stress buildup/relaxation behavior during film growth.

## In-plane texture and epitaxy

Pole figures of the α-$MoO_3${021} Bragg reflections with $2\theta = 27.3^{\circ}$ for α-$MoO_3$ on f-mica show eighteen diffraction features: six singlets and six doublets (Fig. 2a). The residual intensity at other $\phi$ angles was non-zero but weak, indicating in-plane texture in the α-$MoO_3$ film. The singlets and doublets repeat every $60^{\circ}$ and can be described by $\phi_{singlet} = n \times 60^{\circ}$, and doublets are centered at $\phi_{doublet} = \phi_{singlet} + 30^{\circ} = n \times 60^{\circ} + 30^{\circ}$, where n = 0, 1, 2, and 3 (see Fig. 2b). The two spots within each doublet are azimuthally separated by $2\Delta \approx 7.0^{\circ}$, i.e., $\Delta \approx 3.5^{\circ}$ is the angular separation of each doublet spot from the doublet center.

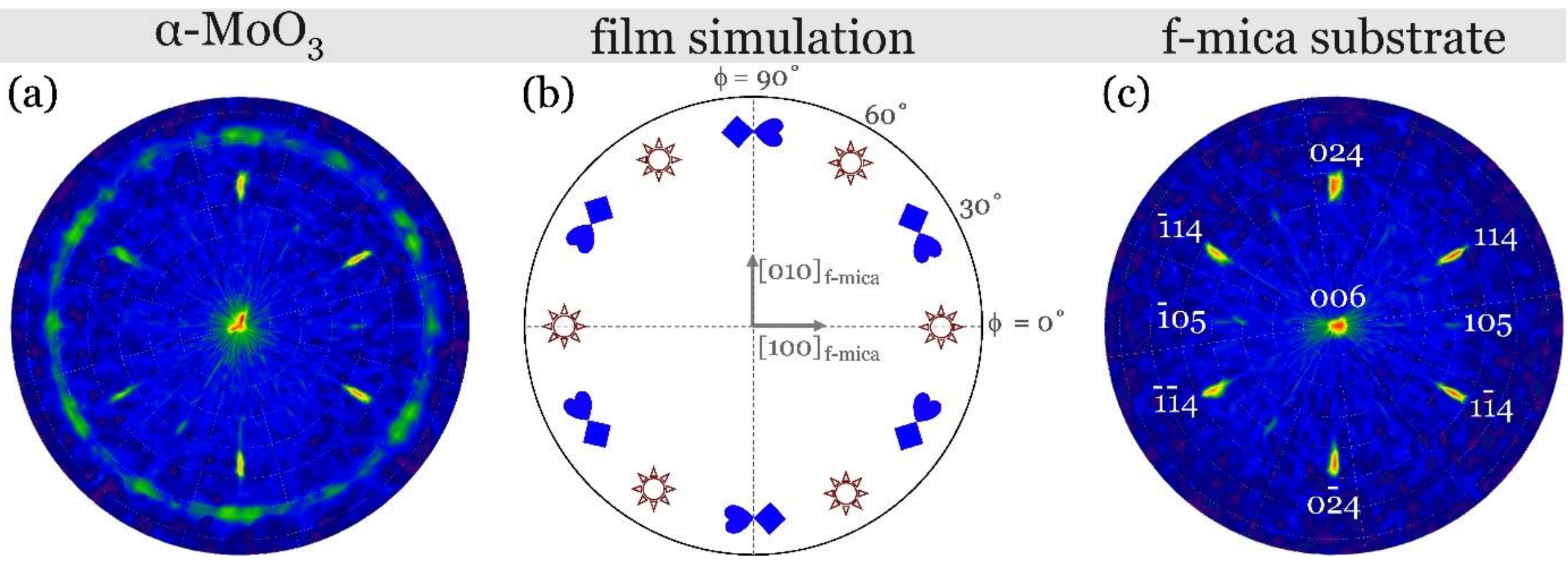

Fig. 2. X-ray pole figure for α-$MoO_3${021} reflections from α-$MoO_3$ epilayers on f-mica (a). A schematic sketch of the α-$MoO_3${021} spots with each symbol connoting a distinct in-plane domain (b). A reference pole figure from a bare f-mica substrate, i.e., without a α-$MoO_3$ film (c).

The $\alpha$-$MoO_3$ doublet centers are radially aligned with {011} and {112} reflections from f-mica (Fig. 2c). The singlets are azimuthally midway between the doublet centers, and are radially aligned with weak f-mica{102} reflections. Since $\alpha$-$MoO_3$(021) has two-fold rotational symmetry about the (001) normal, it is sufficient to describe the $0^{\circ} \leq \phi \leq 90^{\circ}$ sector, with $\phi = 0^{\circ}$ and $90^{\circ}$ pointing to $[100]_{f\text{-}mica}$ and $[010]_{f\text{-}mica}$, respectively.

Two-fold rotations about the (001) normal produce other equivalent doublets and singlets from $\alpha$-$MoO_3$ domains with $\Delta\phi = 60^{\circ}$, consistent with the pseudohexagonal symmetry of f-mica. Thus, $\alpha$-$MoO_3$ crystal growth on f-mica evolves in three in-plane domain orientations. The presence of such in-plane texture together with 0k0 out-of-plane texture shown earlier, indicate $\alpha$-$MoO_3$ epitaxy on f-mica. In contrast, pole figures from $\alpha$-$MoO_3$ films on c-sapphire show continuous rings, indicative of 0k0 only out-of-plane fiber texture (Fig. S2). Thus, f-mica supports $\alpha$-$MoO_3$ epitaxy, c-sapphire only supports out-of-plane fiber texture.

**Proximal atomic correlations across the film/substrate interface**

The interface atomistic structures for the singlet and the doublet configurations were obtained by superimposing orthorhombic $\alpha$-$MoO_3$ and monoclinic f-mica layers with $(010)_{\alpha\text{-}MoO_3} || (001)_{f\text{-}mica}$ at different $[100]_{\alpha\text{-}MoO_3}$ - $[100]_{f\text{-}mica}$ azimuthal angles denoted by $\phi$ (see Fig. 3). For clarity, only K atoms (purple) in f-mica and Mo atoms (green) in $\alpha$-$MoO_3$ are represented in Fig. 3. The detailed structures are shown in supplemental Fig. S3.

For the singlet configuration denoted by $\phi_1 = 0^{\circ}$ (Fig. 3a), the K atoms from f-mica are strongly correlated with and proximal to the Mo atoms in $\alpha$-$MoO_3$ along $[100]_{\alpha\text{-}MoO_3}$, with every fifth atomic row along $[001]_{\alpha\text{-}MoO_3}$ showing near-perfect registry. The two doublet spots denoted by $\phi_2$ and $\phi_3 = 30^{\circ} \pm \Delta$ with $\Delta \approx 3.5^{\circ}$ (Figs. 3b-c) correspond to two non-equivalent interface structures with different cross-interfacial atomic proximities. For $\phi_2 \approx 26.5^{\circ}$, the Mo atoms along α-$MoO_3$[100] are close to K atoms along $[100]_{f\text{-}mica}$, with every third row showing perfect registry (Fig. 3b). For $\phi_3 \approx 33.5^{\circ}$, the Mo atoms are in near-perfect registry every fourth row of K atoms (Fig. 3c). These differences indicate that the doublet spots $\phi_2$ and $\phi_3$ are associated with non-equivalent atomic registries. This non-equivalence is consistent with our depiction of the two spots of every doublet by different symbols in the pole figures (Fig. 2b).

The narrow azimuthal widths of the singlets ($\Delta\phi_1 \approx \pm 1.8^{\circ}$) and the doublet spots ($\Delta\phi_2 \approx \Delta\phi_3 \approx \pm 0.85^{\circ}$) indicate the precise alignments of α-$MoO_3$ crystals with the f-mica substrate for the three interface configurations. The lack of crystal growth for even slightly different $\phi$ angles suggests that the interface crystal configurations for $\phi_1$, $\phi_2$ and $\phi_3$ azimuthal angles correspond to trans-interface atomistic proximity over large in-plane distances that drives vdWE. Such strong dependence of epitaxy on trans-interfacial atomic proximity is consistent with weak short-range secondary bonding specified by $1/r^6$ attractive potentials. Strained primary cross-interface bonding ($1/r$ attractive potential) is energetically precluded at interfaces between two layered materials with strong intra-layer bonding.

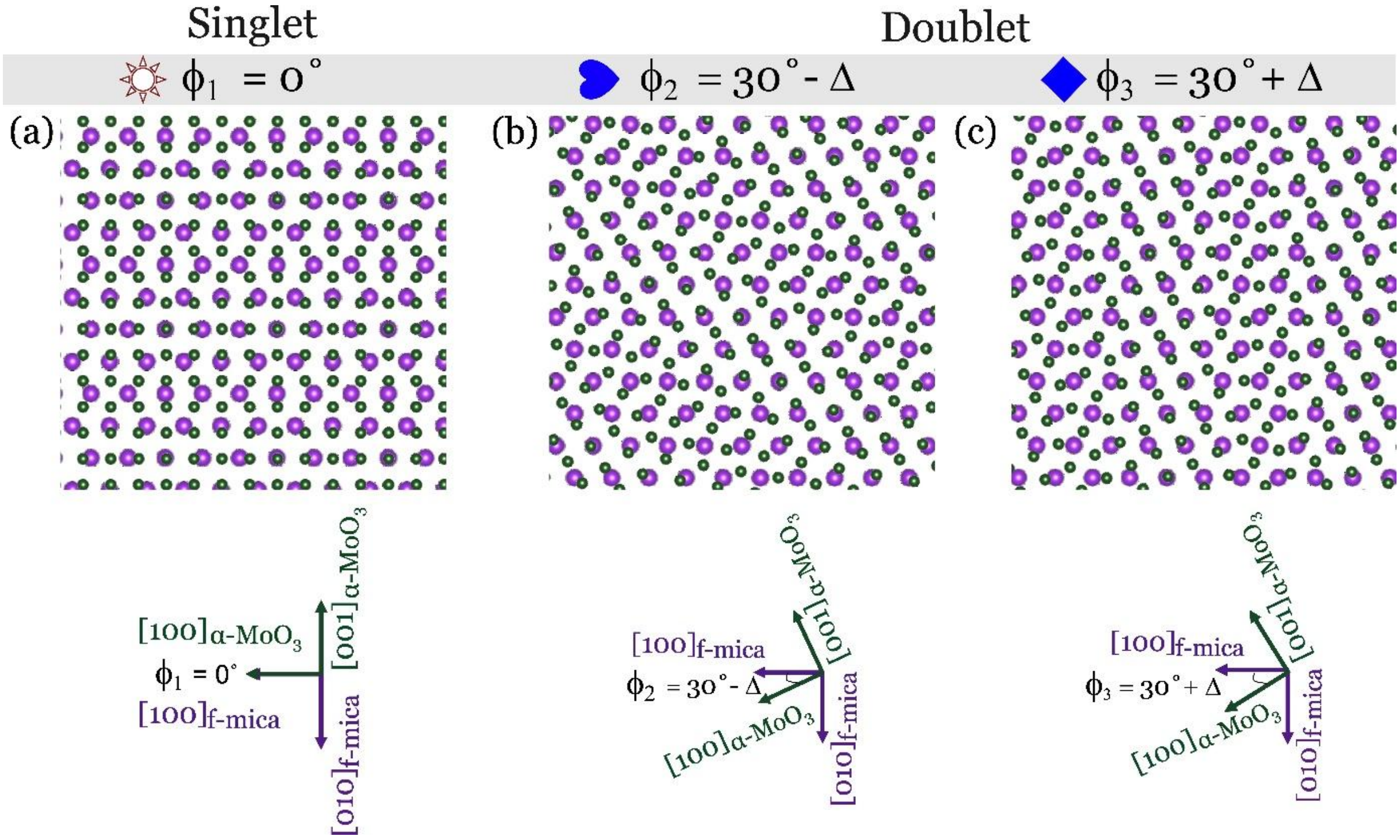

Fig. 3. Interface atomic configurations of α-$MoO_3$ and f-mica corresponding to the α-$MoO_3$ (021) (a)

singlet at $\phi_1 = 0^\circ$, and doublet spots at $\phi_2 = 30^\circ - \Delta$ (b) and $\phi_3 = 30^\circ + \Delta$ (c) , with $\Delta \approx 3.5^\circ$. The $\alpha$-$MoO_3$ structure is represented by Mo ions (green), and the (001) f-mica substrate is represented by K ions (purple). The other elements are not shown to retain visual clarity; see Fig. S3 for detailed structures.

**Energetics and mechanism**

The primacy of vdW bonding at the $\alpha$-$MoO_3$/f-mica interface is supported by *ab initio* computations showing that the in-plane displacement energy for sliding $\alpha$-$MoO_3$ islands on f-mica is comparable to the graphene-graphene glide energy. In contrast, sliding homoepitaxial interfaces of non-layered Mo(110) and TiN(001) requires ≈ 20- to 60-fold greater energies (Fig. S4.1a). For generating sufficient magnitude of vdW interactions to support epitaxy, trans-interface atomistic proximity needs to extend over sufficiently large in-plane distances. We envision vdWE for $\alpha$-$MoO_3$/f-mica interface configurations with energy minima corresponding to maximum trans-interface atomistic proximity.

Our *ab initio* calculations indeed show interface energy minima for three in-plane $\alpha$-$MoO_3$(0k0)/f-mica(001) configurations, namely, $\phi_{th1} \approx 1.5^\circ$ and $\phi_{th2}$ and $\phi_{th3} \approx 33^\circ \pm \Delta_{th}$ with $\Delta_{th} \approx 1.3^\circ$ (Fig. 4a). These angles are close to the singlet at $\phi_1 = 0^\circ$ and doublet spots $\phi_2$ and $\phi_3$ at $30^\circ \pm \Delta$ with $\Delta \approx 3.5^\circ$ from experiment. Such agreement is remarkable because our calculations were for $\approx 0.8 \times \approx 0.7$ $nm^2$ islands with no long-range trans-interface proximity.

Examining larger $\alpha$-$MoO_3$(0k0) slabs (e.g., $11 \times 11$ $nm^2$) on f-mica(001) reveals that the $\phi$ angles of interface energy minima correlate with trans-interface Mo-K separation minima (Fig. S4.2a-b). These results reiterate that in-plane texturing is driven by vdW interactions underpinned by high trans-interface proximity. Even slight deviations from the minima suppress crystal growth due to substantially diminished trans-interface proximity, and hence, decreased vdW interface bonding. Thus, transient $\alpha$-$MoO_3$ nuclei of other orientations, if formed, would likely dissolve and coalesce with low-interface-energy $\alpha$-$MoO_3$(0k0) crystals, yielding the observed texture.

Our simulations reveal Mo-O-K bridges between $\alpha$-$MoO_3$ island edges and f-mica (Figs. 4b-d) that relax back onto the interface during $\alpha$-$MoO_3$ growth. Such transient ionic-covalent interactions likely stabilize fledgling $\alpha$-$MoO_3$ nuclei and diminish with continued $\alpha$-$MoO_3$ crystal growth (Fig. S4. 3) as trans-interface proximity is established over large in-plane distances. Non-transient K-O-Mo bridges are unlikely because they would support film strain, contrary to our experimental results.

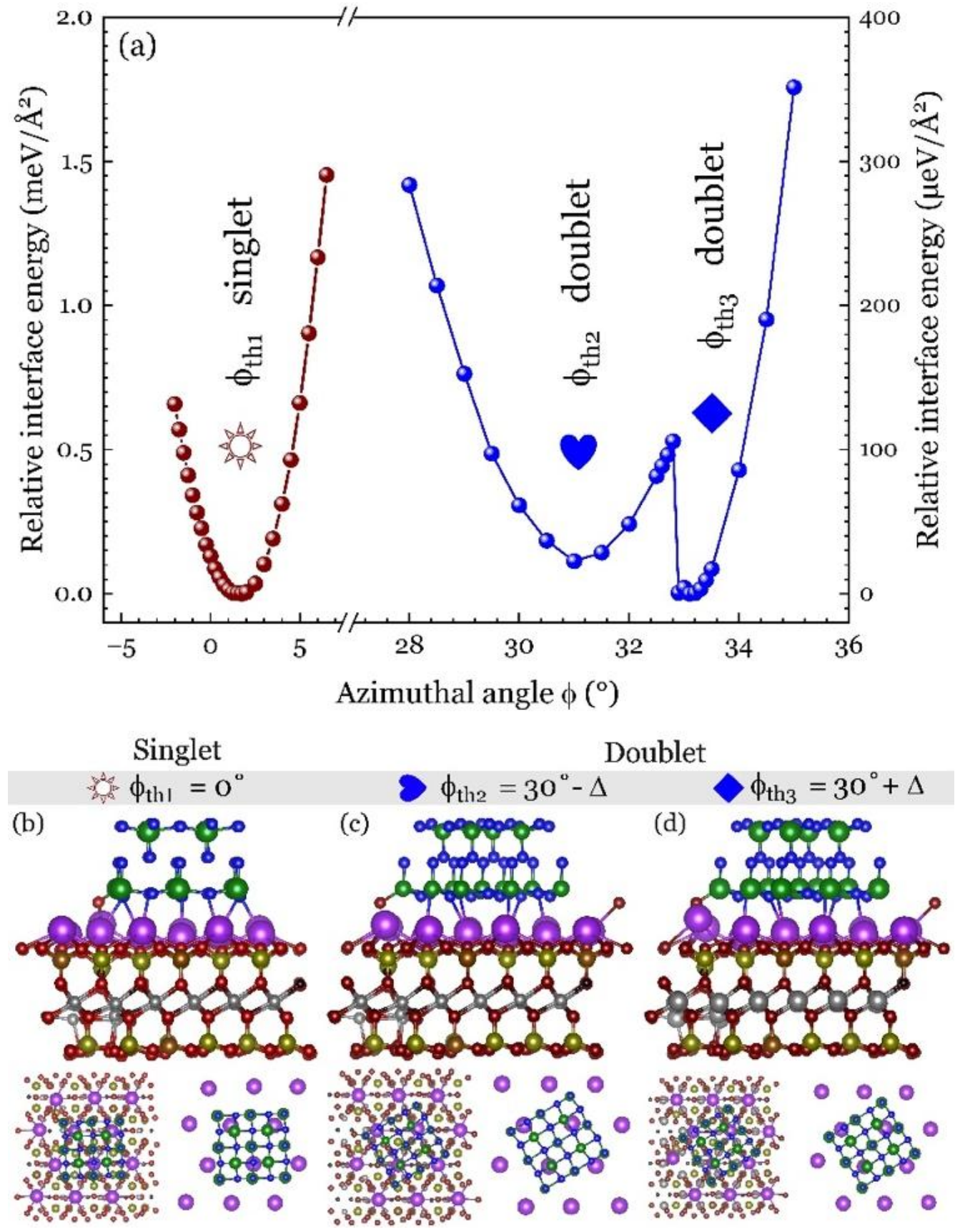

Fig. 4. Computed interface energy plotted versus azimuthal angle $\phi$ between $[100]_{\alpha\text{-}MoO_3}$ - $[100]_{\text{f-mica}}$ in the vicinity of three interface atomistic configurations yielding vdWE, namely, $\phi_{th1} = 1.5^\circ$, and $\phi_{th2}$ and $\phi_{th3} = 30^\circ \pm 3.5$ (a) . Cross-section, plan-view and atomic-registry illustrations of the most stable $\alpha$-$MoO_3$ island on f-mica for interface configurations with minima $\phi_{th1}$ at (b), $\phi_{th2}$ (c), and $\phi_{th3}$ (d). For clarity, only the surface K atoms of f-mica and Mo and O atoms from $\alpha$-$MoO_3$ closest to the substrate are shown. Scheme: O in f-mica = red; O in $\alpha$-$MoO_3$ = blue; K = purple; F = black; Al = brown; Mo = green; Si = yellow-green; Mg = grey.

## Crystal size and interface structure

High angle annular dark-field scanning transmission electron microscopy (HAADF STEM) and selective area electron diffraction (SAED) analyses (Fig. 5) corroborate our X-ray diffraction results and provide visual insights on the vdWE growth of α-$MoO_3$ on f-mica. Cross-section images reveal α-$MoO_3$(0k0) ≥100-nm-wide columnar crystals (Fig. 5a). The uniform contrast and sharp diffraction spots indicate high crystal quality. The seamless continuity of atomically sharp lattice fringes across the film-substrate interface indicates high-quality epitaxy (Fig. 5b), confirmed by SAED patterns showing alignment of α-$MoO_3$ and f-mica spots (Fig. S5). Atomic-resolution images (e.g., Fig. 5c) along α-$MoO_3$ [100] and their fast Fourier transforms (FFTs) indicate atomic arrangement and periodicity consistent with that of orthorhombic α-$MoO_3$.

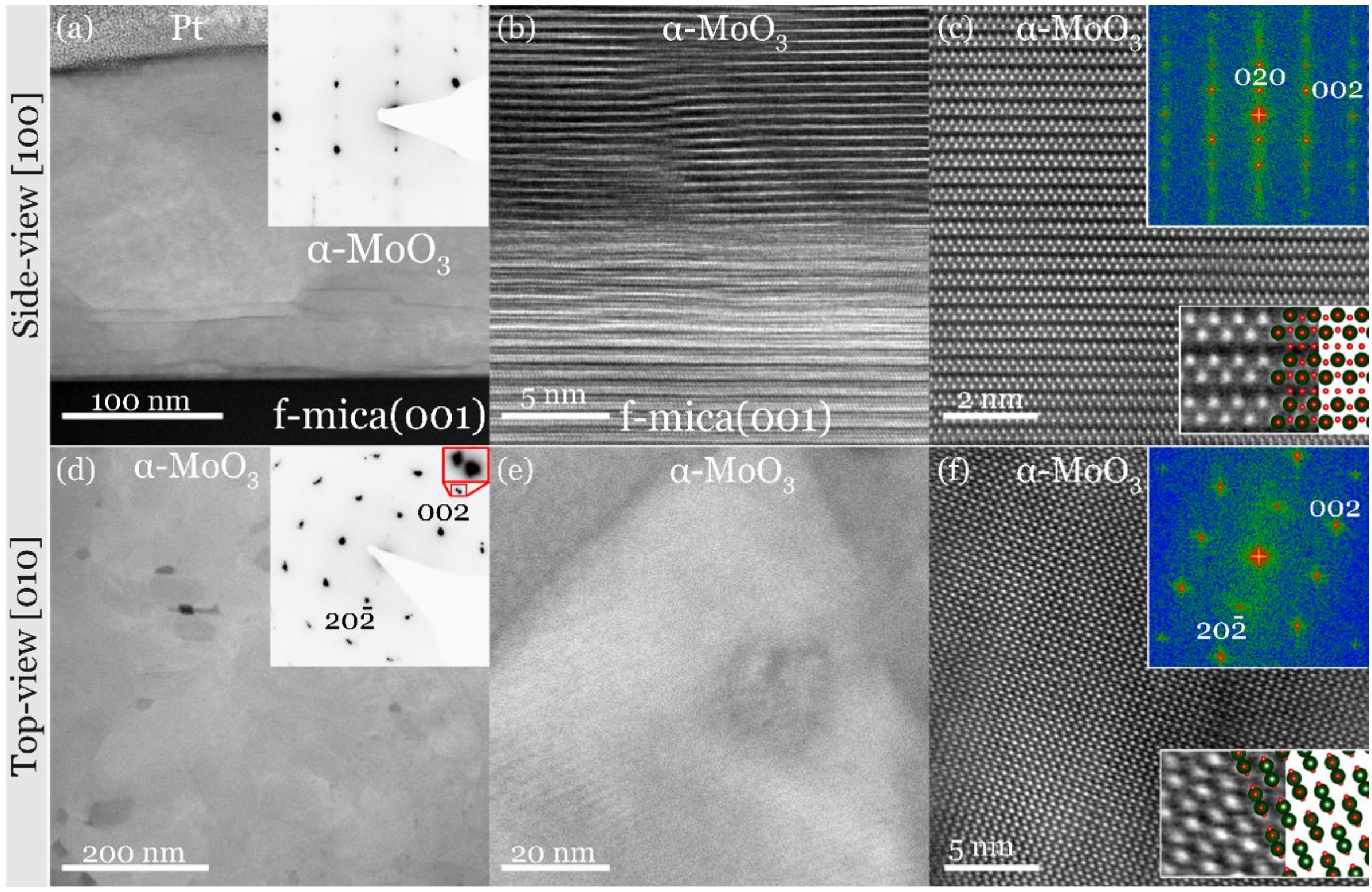

Fig. 5. HAADF STEM images of α-$MoO_3$(0k0) epilayers on f-mica: (a-c) side-view along $MoO_3$ [100] and (d-f) plan-view along $MoO_3$ [010]. (a) A low-magnification overview of the α-$MoO_3$ film with a SAED pattern inset. (b) A high-magnification image capturing the α-$MoO_3$/f-mica interface structure. (c) An atomic-resolution image of a α-$MoO_3$ crystal with insets showing the FFT, and a portion of the image overlaid with the orthorhombic $\alpha$-$MoO_3$ structure. (d) A low-magnification overview of the $\alpha$-$MoO_3$ film, with a SAED pattern (inset) showing doublet diffraction (deep inset). (e) A high-magnification image of a $\alpha$-$MoO_3$ crystal. (f) Atomic resolution image from a $\alpha$-$MoO_3$ crystal. Insets show the FFT, and a portion of the image overlaid with the orthorhombic $\alpha$-$MoO_3$ structure.

Plan-view images confirm an epilayer with densely-packed $\geq$100-nm-wide $\alpha$-$MoO_3$ columnar crystals (Fig. 5d). The occasional presence of dark and grey spots indicate intragranular nanovoids. Such features were not observed in cross-section images, suggesting that the voids may be due to ion beam damage during sample preparation rather than crystal growth. SAED diffraction doublets (Fig. 5d deep inset) indicate multiple in-plane domains (Figs. 5d-e), consistent with our X-ray diffraction results. The $\alpha$-$MoO_3$(010) structure and orientation from HAADF-STEM images and their FFTs are in good agreement with $\alpha$-$MoO_3$ atomic crystal structure (Fig. 5f).

## Summary

We have demonstrated for the first time that vdWE of a layered material such as $\alpha$-$MoO_3$ on a layered substrate f-mica is overarchingly underpinned by proximal cross-interface atomic registry over large in-plane distances. X-diffraction and transmission electron microscopy show that $\alpha$-$MoO_3$ epilayers consist of $>$100-nm-wide columnar crystals with a strong out-of-plane $\alpha$-$MoO_3$(0k0) texture in three sets of non-equivalent in-plane orientations. *Ab initio* computations of $\alpha$-$MoO_3$/f-mica interface energetics confirm that $\alpha$-$MoO_3$ epitaxy is driven primarily by cross-interface vdW interactions, consistent with negligible strain build-up in continuous $MoO_3$ epilayers. The three in-plane $\alpha$-$MoO_3$ domains correspond to energy minima that correlate with maximal cross-interface proximity between Mo atoms $\alpha$-$MoO_3$ and K atoms in f-mica. These definitive first-time atomistic insights on the vdWE mechanism should provide a framework for explaining and predicting vdWE for different film/substrate combinations, and pave ways for designing stress-free and/or stand-alone epitaxial films of layered materials such as α-$MoO_3$ on layered substrates such as f-mica.

**Acknowledgements**

The authors acknowledge funding from the Swedish Government Strategic Research Area in Materials Science on Functional Materials at Linköping University (Faculty Grant SFO-Mat-LiU No. 2009 00971), the Knut and Alice Wallenberg foundation through the Wallenberg Academy Fellows program (KAW-2020.0196), the Swedish Research Council (VR) under Project Nos. 2021-03826, 2025-03680 (P. E.), and 2025-03760 (A. F.), 2024-04996 (G. R.), and the Swedish Energy Agency under project number 52740-1.

This work was partially supported by the Wallenberg Initiative Materials Science for Sustainability (WISE) funded by the Knut and Alice Wallenberg Foundation, and the US NSF grant CMMI 2135725 through the BRITE program, and the Empire State Development's Division of Science, Technology and Innovation Focus Center at RPI (C210117).

All simulations were enabled by resources provided by the National Academic Infrastructure for Supercomputing in Sweden (NAISS) at NSC in Linköping and PDC in Stockholm partially funded by the Swedish Research Council through Grant Agreements No. 2022-06725. DGS gratefully acknowledges financial support from the Swedish Research Council (VR) through Grant Nº VR-2021-04426 & 2025-04266.

VR and Swedish Foundation for Strategic Research are acknowledged for access to ARTEMI, the Swedish National Infrastructure in Advanced Electron Microscopy (2021-00171 and RIF21-0026).

## Experimental details

### *Thin film synthesis*

α-$MoO_3$ films were deposited by pulsed dc reactive magnetron sputter deposition in an ultrahigh vacuum chamber described elsewhere [39]. The depositions were carried out on fluorphlogopite $KMg_3(AlSi_3O_{10})F_2$ (referred henceforth as f-mica) and c-$Al_2O_3$ (c-sapphire) for comparison. The f-mica and c-sapphire substrates were acquired from Continental Trade Sp. z o.o., and Alineason Materials Technology GmbH, respectively.

A 50-mm-diameter 99.99% purity Mo target (Plasmaterials) was sputtered with a 0.33 Pa (2.5 mTorr) $O_2$/Ar plasma generated by magnetrons powered by 150 W with 100 kHz bipolar dc voltage pulse with a 2 μs and duty cycle of 80% to inhibit arcing. The substrate temperature was maintained at $T_{substrate}$ = 400 °C and the oxygen flow ratio was $f_{O2} = O_2/[Ar+O_2]$ =20/58. Analyses of X-ray reflectivity measurements from the films yielded an average deposition rate of 1.7 nm $min^{-1}$. Film thicknesses in the $2.5 \leq t_{film} \leq 160$ nm range were obtained by adjusting the deposition time $t_{deposition}$ [40].

Prior to each deposition, the fresh surface of synthetic f-mica substrate was exposed by mechanical exfoliation using a tape and not cleaned further. The c-sapphire substrate was successively cleaned ultrasonically in acetone and isopropanol for 5 minutes and blow-dried with $N_2$. The substrates were mounted on a rotatable sample holder, and the chamber was pumped to a base pressure of $3\times10^{-6}$ Pa. The synthetic f-mica surfaces were preheated to 400 °C and held for 15 minutes to dispel and minimize adsorbed water, carbon dioxide and hydrocarbons. The Mo target was sputter cleaned in an Ar plasma (without oxygen) for two minutes with the substrates covered by a shutter.

***Characterization***

Out-of-plane texture and stress evolution

Bragg Brentano θ-2θ X-ray diffraction (XRD) scans were acquired using a PANalytical X'Pert PRO diffractometer system equipped with a Cu Kα (λ = 1.54 Å) source operated at 45 kV and 40 mA and Ni screen to filter $CuK_{\beta}$. The incident beam passed through 0.5° divergence and anti-scatter slits, and the diffracted beam included a 5.0 mm anti-scatter slit and 0.04-rad Soller slits. A PANanalytical Empyrean diffractometer was used for X-ray reflectivity (XRR) measurements for film thickness determination using hybrid mirror with a 0.5° divergence slit and a 0.125° divergence slit.

XRD pole figure measurements were performed with an Malvern Panalytical Empyrean X-ray diffractometer with zero-setpoint crossed-slit primary optics and a 0.27° parallel plate collimator in the secondary optics. Data was collected for 1.0 s/step at 2.5° steps for both tilt $0° \leq \omega \leq 85°$ and azimuthal rotation $0° \leq \phi \leq 360°$ angles. CaRIne Crystallography v3.1®software was used for pole figure simulations and crystallography analyses. Vesta (Visualization for Electronic and Structural Analysis) Ver.3.5.8 was used to create schematic representations of the atomic arrangement.

***Morphology and atomic structure determination***

Scanning electron microscopy (SEM) was performed in an SEM Leo 1550 Gemini (Zeiss) operated with an acceleration voltage of 2 kV and in-lens detector. Cross-sectional and plan-view TEM specimens were prepared using a Helios 5 UC DualBeam FIB-SEM system (Thermo Fisher Scientific) via an *in situ* lift-out procedure. Subsequent TEM and high-angle annular dark-field scanning TEM (HAADF-STEM) analyses were carried using a double Cs-corrected $Titan^3$ 60-300, operated at 300 kV.

***Computational details***

Density functional theory (DFT) calculations are carried out using VASP implemented with the projector augmented-wave method [41,42]. The calculations employ Γ-point sampling of the reciprocal space while the electronic exchange and correlation effects are described by the Perdew-Burke-Ernzerhof (PBE) approximation [43]. van der Waals corrections to DFT energies are modelled according to Grimme [44].

DFT calculations provide insights on energetically preferred configurations of $\alpha$-$MoO_3$ islands on f-mica. First, the structures of pristine 3D-periodic $\alpha$-$MoO_3$ and f-mica crystal models are fully optimized by conjugate gradient relaxation of atomic positions, supercell shape and volume. The convergence criteria on energies and forces are set to $10^{-5}$ eV/supercell and $10^{-2}$ eV/Å. Supercells of $\alpha$-$MoO_3$ and f-mica contain 144 and 240 atoms (12 K, 36 Mg, 120 O, 24 F, 12 Al, and 36 Si), respectively, with randomly distributed Al and Si atoms. The bulk supercells sizes are ~ $1.2 \times 1.4 \times 1.1$ $nm^3$ ($\alpha$-$MoO_3$) and ~ $1.6 \times 1.8 \times 1.0$ $nm^3$ (f-mica). DFT calculations were then used to determine the interface energies between a $\alpha$-$MoO_3$ island and the f-mica as a function of the island position and rotation about the azimuthal angle $\phi$.

The smaller and larger square-like shape $\alpha$-$MoO_3$ island models (50 and 72 atoms in Fig. 4 and Fig. S4.3, respectively) have in-plane edges parallel to [100] and [001] crystal axes, and a size of ~ $1 \times 1 \times 0.5$ $nm^3$. The $\alpha$-$MoO_3$/f-mica interface structure is optimized in a sequence of steps. We begin by relaxing a f-mica surface slab containing 24 K, 72 Al, 72 Si, 48 H, and 288 O atoms. The $\alpha$-$MoO_3$ island is constructed by cleaving the $\alpha$-$MoO_3$ bulk crystal previously relaxed by DFT. It is positioned approximately 2 Å above the relaxed f-mica surface, with the island and f-mica [100] directions aligned along the Cartesian x-axis and the island $[001]_{\alpha\text{-}MoO_3}$ aligned with $[010]_{f\text{-}mica}$ along the y-axis. The $\alpha$-$MoO_3$/f-mica simulation box contains a 14-Å vacuum layer that separates supercell replicas along the surface normal direction. At this stage, all atomic coordinates within the $\alpha$-$MoO_3$ island are kept fixed while the f-mica atoms are fully relaxed. This allows the substrate to adjust to the presence of the island and yields the locally most stable interfacial configuration for that fixed island geometry.

After identifying the most stable interface geometry, we map the interface energy $E_{interface}$ as a function of the lateral displacement of the $\alpha$-$MoO_3$ island along x and y. The island is shifted in 0.1 Å increments even as the internal atomic coordinates remain fixed. For each lateral position, the f-mica slab is relaxed except for its bottom oxygen layer, which is kept frozen. An analogous procedure is used to determine how $E_{interface}$ varies with the azimuthal rotation angle $\phi$ between the island and substrate [100] axes.

Motivated by experimental observations, we first probe $E_{interface}$ upon small changes around $\phi \approx 0^\circ$ reminiscent of the experimental singlet $\phi_1$ (Fig. 2). For each $\phi$, the island atoms remain fixed, and the f-mica slab is fully relaxed except for its bottom oxygen layer. We then impose a $\phi = 30^\circ$ rotation of the island about its vertical [010] axis and keep the island rigid, while all underlying f-mica atoms are relaxed to accommodate the rotated island. Starting from this reference optimized configuration, the island is subsequently rotated by small positive and negative angular increments around $\phi = 30^\circ$. The interface energy at each displacement is computed as $E_{interface}(x,y \mid \phi) = [E_{\alpha\text{-}MoO_3/f\text{-}mica}(x,y \mid \phi) - E_{\alpha\text{-}MoO_3\text{-}island} - E_{f\text{-}mica}]/A_{\alpha\text{-}MoO_3\text{-}island}$, where $E_{\alpha\text{-}MoO3/f\text{-}mica}(x,y \mid \phi)$ is the total energy of the combined system with the island positioned at (x,y) with a rotation angle $\phi$, $E_{\alpha\text{-}MoO_3\text{-}island}$ and $E_{f\text{-}mica}$ are energies of the isolated $MoO_3$ island and isolated f-mica, respectively, and $A_{\alpha\text{-}MoO_3\text{-}island}$ is the in-plane area of the $\alpha$-$MoO_3$ island.

## Author Declarations

***Conflict of Interest*** - The authors have no conflicts to disclose.

## Author Contributions

**Faezeh A. F. Lahiji**: Project administration (lead), Conceptualization (lead); Formal analysis (lead); Data curation (lead); Investigation (lead); Methodology (lead); Validation (lead); Visualization (lead); Writing-original draft (lead); Writing-review & editing (lead). **Davide G. Sangiovanni:** Data curation (supporting); Methodology (supporting); Validation (supporting); Visualization (supporting); Writing-review & editing (supporting). **Biplab Paul:** Data curation (supporting); Validation (supporting); Visualization (supporting); Writing-review & editing (supporting). **Justinas Palisaitis:** Data curation (supporting); Methodology (supporting); Validation (supporting); Visualization (supporting); Writing-review & editing (supporting). **Per O. Å. Persson:** Data curation (supporting); Methodology (supporting); Validation (supporting); Visualization (supporting); Writing-review & editing (supporting). **Arnaud le Febvrier:** Data curation (supporting); Methodology (supporting); Validation (supporting); Visualization (supporting); Writing-review & editing (supporting). **Ganpati Ramanath:** Data curation (equal); Methodology (equal); Validation (equal); Visualization (equal); Writing-review & editing (equal). **Per Eklund:** Conceptualization (equal); Data curation (equal); Funding acquisition (equal); Investigation (equal); Methodology (equal); Project administration (equal); Resources (equal); Supervision (equal); Validation (equal); Visualization (equal); Writing-review & editing (equal).

## Data Availability

The data collected in this study will be made available on Zenodo after acceptance.